# Journal of Volcanology and Geothermal Research

# Correlating hydrothermal system dynamics and eruptive activity – A case-study of Piton de la Fournaise volcano, La Réunion

Guillaume Mauri [a,*], Ginette Saracco [b], Philippe Labazuy[c], Glyn Williams-Jones[d]

[a] Centre for Hydrogeology and Geothermics (CHYN), University of Neuchâtel, Emile-Argand 11, CH-2000 Neuchâtel, Switzerland
[b] CNRS-UMR7330 CEREGE, Aix-Marseille-Université, AMU, CdF, IRD, Equipe Modélisation, Europole de l'Arbois, BP 80, 13545 Aix-en-Provence Cedex 4, France
[c] Université Clermont Auvergne, CNRS, IRD, OPGC, Laboratoire Magmas et Volcans, F-63000 Clermont-Ferrand, France
[d] Centre for Natural Hazards Research, Department of Earth Sciences, Simon Fraser University, 8888 University Drive, V5A 1S6, Burnaby, BC, Canada

Article

A b s t r a c t

Piton de la Fournaise volcano, La Réunion Island, is a basaltic shield volcano which underwent an intense cycle of eruptive activity between 1998 and 2008. Self-potential and other geophysical investigations of the volcano have shown the existence of a well-established hydrothermal system within the summit cone. The present study investigates the relationship between changes in the hydrothermal system and eruptive activity at the summit cone of Piton de la Fournaise. Here, we consider the depth of the hydrothermal activity section to be the area where the hydrothermal flow is the most intense along its path. Ten complete-loop self-potential surveys have been analyzed through multi-scale wavelet tomography (MWT) to characterize depth variations of the hydrothermal flow between 1993 and 2008. Our MWT models strongly support the existence of six main hydrothermal flow pathways associated with the main edifice structure. Each of these pathways is part of the main hydrothermal system and is connected to the main hydrothermal reservoir at depth. In both 2006 and 2008, around Dolomieu crater, based on our results, the hydrothermal activity sections are located between 2300 and 2500 m a.s.l., which correlate well with the elevation of the observed fumarole belt within the post-2007-collapse crater wall. Our results show that the depths of the local hydrothermal activity sections change substantially over the investigated period. Vertical displacement of the main potential generation area, associated with these hydrothermal activity sections, is observed on the order of several hundred meters at the transition between the period of quiescence (1993–1997) and the resumption of eruptive activity in 1998 and 2007, respectively. From 1999 to March 2008, the hydrothermal system was consistently located at relatively shallow depths. By quantitatively determining the vertical displacement of hydrothermal fluids over 16 years, we identify a significant link between hydrothermal system and magmatic activity. Hydrothermal fluids depth below the surface is an indicator of the activity level (pressurization/depressurization of the volcano) within the shallow magmatic systems. Thus, when used in conjunction with long term volcano monitoring, this approach can contribute substantially to detection of the precursory signals of changes in volcanic activity.

## 1. Introduction

Since the 1980s Piton de la Fournaise volcano, has been well studied using a wide range of geological and geophysical data, leading to the development of a comprehensive model of the volcanic edifice and of the hydrogeologicalsystem within the summit (Lénat & Bachelery1990; Lénat et al 2000; Peltier et al 2012, 2009; Gailler & Lénat 2012; Lénat et al 2012a,b) along with new insights on the high level of eruptive activity from 1998 to 2008 (Peltier et al., 2005, 2006).

In the summit part of the volcanic edifice, the hydrogeological system consists of system centred ata well-developed hydrothermal depth beneath the craters and of lateral outflows toward the ocean (Fig. 1) (Join et al., 2005; Lénat et al.,2012a; Barde-Cabusson et al., 2012 and references therein). While this hydrothermal system has no permanent manifestation at the surface, its impact on the dynamics of the volcanic activity has been shown to be non-negligible (Saracco et al, 2004; Gouhier and Coppola, 2011; Lénat et al., 2012a, b; Barde-Cabusson et al., 2012; Peltier et al., 2012).

Previous models, as well as field observations following the large collapse of Dolomieu crater during the 2007 eruption, suggest that hydrothermal fluids and magma use similar pathways to reach the surface (Staudacher et al 2009; Staudacher 2010). The hydrothermal system is believed to consist of one large system at depth which splits into several main hydrothermal flow paths that extend toward the surface along complex pathways (*Barde-Cabusson et al and Peltier et al, 2012*).The repeated collapses of the summit craters have led to the formation of a heterogeneous highly permeable and vertical collapse structure delimited by faults, which acts as a preferential drain for magmatic intrusion and hydrothermal fluid circulation (*Saracco et al 2004, Lénat et al 2012a*).

* Corresponding author.
E-mail addresses: guillaume.mauri@unine.ch (G. Mauri), saracco@cerege.fr (G. Saracco), P.Labazuy@opgc.fr (P Labazuy), glynwj@sfu.ca (G Williams-Jones)

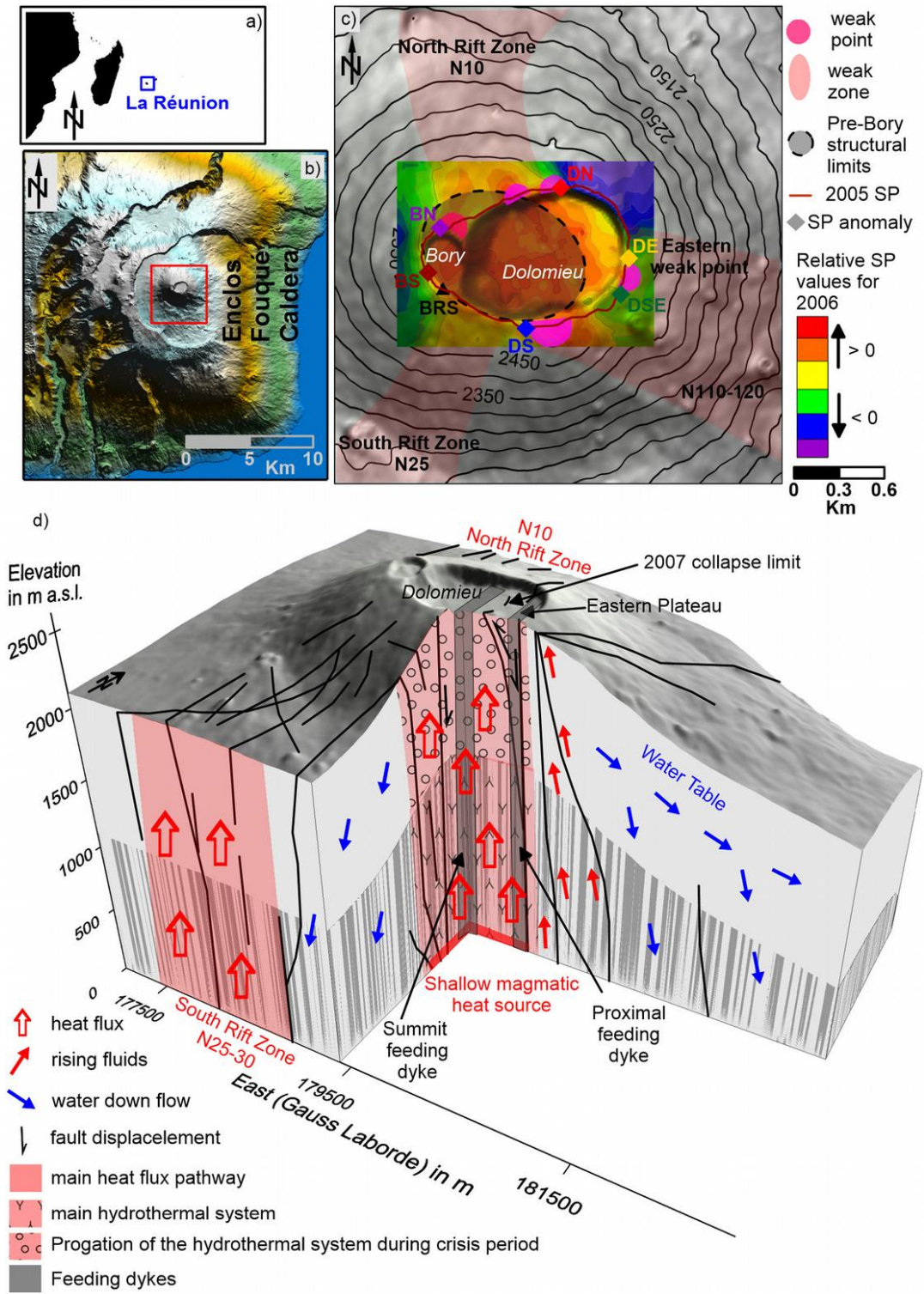

Fig.1. Piton de la Fournaise volcano (PdFV) La Réunion. a) Geographic location of La Réunion Island. b) Shaded relief map of PdFV, located in the South-East part of La Réunion Island. The red rectangle represents the location of the summit cone in the Enclos Fouqué caldera. c) The summit cone of PdFV with the craters: Bory and Dolomieu. The extent and location of structural limits (shaded areas); self-potential anomalies (diamonds) are defined based on the literature (*Michon et al 2009a; Peltier et al 2009a; Barde-Cabusson et al. 2012*). The red loop represents new data from our study with the self-potential profile surveyed during the eruption of March 2005. BRS is the self-potential Bory reference station. The main SP anomalies are named based on their geographical location: BS is Bory South; DS is Dolomieu South; DSE is Dolomieu South-East; DE is Dolomieu East; DN is Dolomieu North; BN is Bory North. d) 3D model of the hydrogeological structures as proposed *in Lénat et al 2012a, Barde-Cabusson et al 2012* which incorporate previous models from the literature (*Lénat et al 2000; Michon et al 2009a*). The 1000m a.s.l. limit corresponds to the upper border of the main hydrothermal system. Orientation of feeding dyke is based on work of *Peltier et al. (2009)*. The width of the dyke is not to scale.

While many uncertainties still exist regarding the exact geometry of this High permeability zone, it appears that its shape, projected at the surface Is similar in size to the 2007 caldera roof collapse. The pit crater structure located on the top of the summit cone is considered to act as a drain to these rising hydrothermal fluids (*Lénat et al 2012a*). Given this structural relationship, our study seeks to determine if there is any correlation between changes in the hydrothermal system and the eruptive activity. This may contribute to the detection of signals precursory to changes in volcanic activity of Piton de la Fournaise.

Of the extensive geophysical data sets collected, frequent self-potential (SP) surveys have constrained the surface-projected distribution of the hydrothermal system on Piton de la Fournaise (*Malengreau et al 1994; Lénat et al 2000, 2012a; Barde-Cabusson et al 2012*). Previous work on complex and real wavelet analyzes and dipolar probability tomography of this self-potential data (1993, 1998 and 1999 surveys) showed the presence of geophysical discontinuities at depth linked with subsurface water flow (*Saracco et al 2004*) and highlighted a possible correlation between the estimated depth of hydrothermal circulation and volcanic activity. However, further work was needed to investigate the possible control of faults and areas of high permeability on the location of the main active sections in the hydrothermal system. In this study we build on recent advances in both the understanding of the hydrothermal system (*Lénat et al 2012a; Barde-Cabusson et al 2012*) and wavelet tomography on Piton de la Fournaise and other volcanoes (e.g. *Saracco et al 2004; Mauri et al 2010, 2012; Caudron et al 2016*).We present unpublished SP data collected during the March 2005 eruption, which along with published data (*Malengreau et al., 1994; Lénat et al 2000; Saracco et al 2004; Levieux 2004; Lénat et al 2012a; Barde-Cabusson et al 2012*), are analyzed using « Multi-Scale Wavelet Tomography » to investigate the relationship between the volcanic activity from 1993 to 2008 and variations in the depth of the hydrothermal system. Here we define 1) the electrical generation type as the type of processes that generate the measured SP value on the surface and 2) the potential generation area as the area at depth where the potential generation is taking place.

## 2. Geological setting

Piton de la Fournaise is a basaltic shield volcano, located in the Indian ocean, which over the last 500,000 years, has been shaped by several caldera and flank collapses (*Fig. 1*a, b) (*Bachèlery 1981; Lénat & Bachelery 1990; Labazuy 1996; Merle & Lénat 2003; Carter et al 2007; Oehler et al 2008; Michon et al 2009a,b; Brenguier et al, 2012; Got et al 2013; Sens-Schönfeldera et al 2014; Bachèlery et al 2016; Michon et al 2016*). Inside the Enclos Fouqué caldera, a dense fault system connects the northern and southern rift zones and converges toward the summit cone (*Fig. 1*c) (*Lénat and Bachelery, 1990; Carter et al 2007; Lénat et al 2012b* and reference therein). The eastern side of the summit cone is affected by conjugate faults, which are controlled by the eastward spreading of the east flank of the volcano *(Froger et al., 2004; Carter et al., 2007; Michon et al., 2009a)*.

Recent volcanic activity is principally focused on a large summit cone formed within the Enclos Fouqué caldera (formed 4200 years ago),which hosts two main craters (*Fig. 1* b,c), Bory and Dolomieu, the latter formed in 1930 (*Peltier et al 2012* and references therein). During the 2007 eruption, Dolomieu crater collapsed by over 340 m and has since partially been refilled by lava flows (*Staudacher 2010; Peltier et al 2012*). The internal structure of the summit cone consists of alternating lava flows and scoria layers, which are intruded by a well-developed dyke network (*Fontaine et al., 2002; Battaglia et al., 2005; Peltier et al 2005, 2007, 2009, 2012; Brenguier et al 2007*). The magma storage system is formed by a complex structure of dikes and sills located within and immediately beneath the summit cone. This has been imaged by 3D seismic tomography (*Brenguier et al., 2007*) and inverse deformation modeling (*Peltier et al 2005, 2006, 2007, 2009*). These studies show that the plumbing system is dominated by three distinct dyke networks: the summit, proximal and distal dyke feeder networks. The summit intrusion pathway is controlled by the ring fault structure below Dolomieu crater, while the proximal pathway is located in the eastern part of the summit cone (*Fig. 1*d). Both feeder systems are thought to originate near the top of the shallow magma reservoir (located at ~0 *m* a.s.l.) *(Peltier et al 2009; Staudacher 2010; Lénat et al 2012a, b)*. The distal dyke feeder structures supply magma to the rift zone away from the summit (*Peltier et al., 2009*).

Piton de la Fournaise is very active with 67 eruptions from 1980 to 2018 and 30 effusive eruptions during the 1998–2008 period of this study.The activity reached its highest point with the collapse of Dolomieu crater during the 2007 eruptions. Between 1997 and 2005, the strongest volcanic activity occurred in 2003, with 4 eruptions and 1 shallow intrusion (*Peltier et al 2005, 2009, 2012; Roult et al 2012*).

The time from 1993 to 2008 can be separated into three distinct periods. The 1993–1997 period includes little to no observable magmatic activity, with only one intrusion in 1996 (*Roult et al 2012*). The 1998–2001 period marks the resumption of volcanic activity with 2 to 3 effusive eruptions per year, which took place on the proximal zone around the summit cone, along the N25–30 and N120 rift zones (Fig 1) (*Peltier et al 2009; Massin et al 2011; Lénat et al 2012; Roult et al 2012*). Starting in 2000, the eruption cycles initiated with summit eruptions and during the 2001–2008 period, volcanic activity increased from 3 to 5 eruptions per year. Each of these eruptions started at the top of the summit cone with magma rising along or through the high permeability structure after which eruptive activity migrated the lower elevations. The proximal dyke network is responsible for transporting magma toward more distal areas within the Enclos Fouqué caldera (*Peltier et al 2012*).

## 3. Hydrothermal system of Piton de la Fournaise volcano

Volcanic hydrothermal systems are complex with a wide range In shape, intensity and chemical composition (e.g. *Osinski et al 2001; Merlani et al. 2001; Chiodini et al. 2005*). Hydrothermal fluids consist of one or two phases (liquid and/or gas), dominated by meteoric and magmatic fluids, and generally develop into convective fluid cells which shape the hydrothermal system (e.g. *Osinski et al. 2001; Finizola et al. 2002; Antoine et al. 2009*).

Piton de la Fournaise hosts a well-established hydrothermal system, which is characterized by strong self-potential anomalies, up to several hundred mV (*Fig. 2*) (*Malengreau et al 1994; Lénat et al 2000; Saracco et al 2004; Levieux 2004; Lénat et al 2012a; Barde-Cabusson et al, 2012*). This hydrothermal system consists of one extended hydrothermal system, in which several main flow paths spread throughout the summit cone from the main hydrothermal reservoir (*Saracco et al, 2004*).In this study,we define hydrothermal activity sections to be the section along a main hydrothermal flow path, where hydrothermal activity is the strongest and which is responsible for the SP anomalies measured on the surface. The hydrothermal system appears to have a zone of water recharge on the eastern side of the summit cone (the lowest part of the summit rim), which is characterized by a strong and persistent negative self-potential anomaly over the 16 years of the study (*Fig.2*). The limit between the recharge zone and the rising hydrothermal fluids is believed to pass through the eastern part of Dolomieu crater, which coincides with the eastern border of the collapsed structure (*Barde-Cabusson et al 2012*). The full extent of the main hydrothermal reservoir is still not completely understood (*Lénat et al 2012a; Barde-Cabusson et al* 2012). The bottom of the hydrothermal reservoir Is likely the top of the shallow magma reservoir, while the top of the hydrothermal reservoir reaches the upper part of the summit cone. Following the 2007-collapse event, the fresh crater walls revealed evidence of active hydrothermally-altered zones including a belt of persistent fumaroles at 2350 m around the Dolomieu crater wall (*Peltier et al 2012*).The lateral extent of the hydrothermal system appears to be slightly wider than the collapsed structure. It is known that several dominant hydrothermal flow paths rise along faults, fractures and porous medium and preferentially along the faulted cylinder shape that defines the high permeability structure (Fig.1) (*Saracco et al 2004; Lénat et al 2012a; Barde-Cabusson et al 2012*). The location where the active section of hydrothermal fluids is strongest will depend on the flow pressure, which itself depends on several parame-. ters. The main parameters are the temperature of the boiling fluids, the proportions of the fluid phases and the physical properties of the host rocks (e.g. porosity, permeability). This assumption is valid as previous studies on the propagation of fluid in saturated rock have shown the significance of fluid temperature and pressure (*Natale & Salusti 1996, Merlani et al. 2001; Antoine et al. 2009, 2017*).

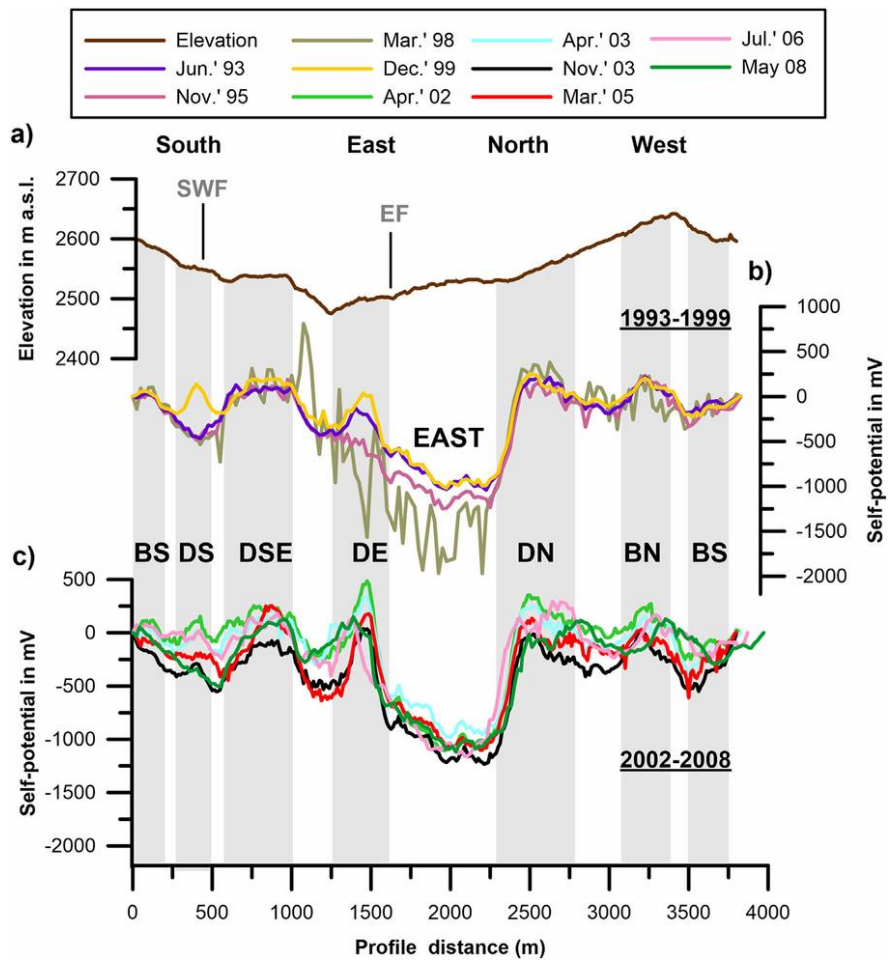

Fig. 2. The 10 self-potential profiles used in this study from 1993 to 2008. a) Topographic elevation. SWF is the south-west fault and EF is the east fault. b) Self-potential profiles from 1993 to 1999 showing the location of the hydrothermal zones (BS, DS, DSE, DE, DN, BN and BS). EAST is the eastern water recharge zone.  c) Self-potential profiles between 1995 and 2008.

Based on field observations, modeling of the magmatic plumbing system (*Peltier et al 2005, 2006, 2009; Brenguier et al 2007; Michon et al 2009b; Lénat et al 2012a; Roult et al 2012*), self-potential analyzes (*Malengreau et al 1994; Lénat et al 2000; Saracco et al. 2004; Barde-Cabusson et al 2012; Antoine et al 2009*), and our understanding of hydrothermal systems (*Osinski et al 2001; Merlani et al 2001; Chiodini et al 2005; Linde et al 2007*), it is reasonable to make the following assumptions regarding the hydrothermal system of Piton de la Fournaise: 1) the hydrothermal fluids are heated and recharged by hot gas and magmatic water from the shallow magmatic reservoir (near sea level, Fig. 1d);   2) the hydrothermal system should vary accordingly with the level of magmatic activity,  and   3) over time, hydrothermal processes may generate hydraulic fracturing or boiling induced decompression within the shallow magmatic reservoir (*Lénat et al 2012a;  Barde-Cabusson et al 2012;  Antoine et al.2017*).

To date, previous models of the hydrothermal system and location of self-potential generation areas suggest that they are a function of the level of magmatic activity (e.g. *Lénat et al 2012a*). Evidence of current or past hydrothermal activity can be observed in surface exposures following the 2007 collapse event. In the case of the Dolomieu collapse in 2007, the work of *Staudacher (2010)*  and *Peltier et al. (2012)* describe the presence of outcrops around the crater at an elevation of 2340 m a.s.l. showing signs of hydrothermal activity, such as high temperatures (100–200 °C), wet surfaces and persistent fumaroles. In addition, previous studies suggest that both hydrothermal fluids and magmatic intrusions follow the same pathway through the collapsed high permeability structure (*Staudacher 2010; Lénat et al 2012a; Barde-Cabusson et al. 2012*). This link was confirmed by the post-2007-collapse crater wall, on which hydrothermal alteration areas follow the main fault structure orientation, pit-crater and injected dike (*Peltier et al 2012*).

During periods of quiescence,  such as between 1992 and 1995, the level of the top of hydrothermal circulation is unknown. However, it is believed to be at least at 1000 m below the topographic surface, corresponding to the upper limit of a low resistivity body assumed to represent the main hydrothermal zone (*Lénat et al., 2000, 2012a*).

Below Dolomieu crater, this limit would thus be around 1300 to 1500 m a.s.l.

This is also supported by other geophysical models (*Lénat et al., 2012a* and references therein). Since the 1998 eruption, both eruptions and intrusions were preceded by earthquake swarms located within the collapsed structure beneath Dolomieu crater from 500 m below the summit to sea level. This seismogenic area corresponds to a low resistivity layer forming the base of the collapsed structure (red surface with "Y" symbol on Fig.1d) extending to a few hundred meters above it. The low resistivity layer is thought to delimit the main hydrothermal reservoir (from sea level to about 1500 m a.s.l.), while the main structural faults cut through the entire collapsed high permeability structure connecting the shallow hydrothermal and magmatic reservoirs to the surface (Lénat et al 2012a). The existence of these faults is defined at the surface by the collapse structures of the 2007 eruption, as well as by the presence of weak points (pit craters) and old crater limits (Fig 1c, 1d). At depth, the faults are defined through the seismic swarm patterns (Lénat et al 2012a) and most seismogenic areas are associated with faults cutting through the western part of Dolomieu crater. The faults act as preferential pathways for the hydrothermal fluids. Therefore, based on the level of hydrothermal activity (during or outside periods of crisis), the depth of the hydrothermal fluids may change overtime (Saracco et al., 2004; Lénat et al., 2012a).

## 4. Methods

### 4.1. Self-potential method

Self-potential (SP) is a passive electrical method that measures the electrical potential between two points, which represent the natural electrical current present in the ground (e.g. Ewing 1939; Poldini 1938; Corwin & Hoover 1979; Ishido & Mizutani 1981; Zlotnicki et al 1994; Zlotnicki & Nishida 2003; Hase et al 2003; Lénat 2007; Aizawa et al 2008; Jouniaux et al 2009; Antoine et al 2017). SP surveys are commonly made using two unpolarizable copper electrodes, consisting of a copper rod in a saturated copper-sulphate solution, connected to an insulated wire cable and a high impedance (100 $M\Omega$) multimeter (e.g. Corwin and Hoover 1979; Ishido and Mizutani 1981; Malengrau et al 1994; Aubert & Atangana 1996; Finizola et al 2002; Hase et al 2003; Aizawa et al 2008; Barde-Cabusson et al 2012; Mauri et al 2012) and more recently with unpolarisable lead-lead electrodes (Petiau-type, lead rod in a lead/lead-chloride solution and sodium-chloride solution;  e.g. Antoine et al., 2009, 2017).

The measured electrical potential corresponds to the electrical field, which is typically the sum of the different types of electrical potential generation processes hereafter called the electrical generation type (Corwin and Hoover, 1979; Zlotnicki et al 1994; Jouniaux et al 2009).

On volcanoes, the three main electrical generation types are  ever, the electrokinetic effect, thermo-electric effect and effect of ground resistivity heterogeneity. On active volcanoes, it is thought that the electrokinetic effect is the main electrical generation process (i.e., Ishido and Mizutani, 1981; Zlotnicki et al.1994; Yasukawa et al 2003). A summary of each type of electrical generation process is presented in Appendix A, while more detailed descriptions of SP generation can be found in the literature (i.e., Corwin and Hoover, 1979; Zlotnicki et al., 1994; Ishido and Mizutani, 1981; Lénat 2007; Jouniaux et al.2009 and references therein).

In the case of Piton de la Fournaise volcano, previous studies have demonstrated that the electrokinetic effect is the main type of electrical generation and associated with groundwater flow displacement through a porous or fractured medium (Saracco et al  2004; Lénat 2007; Antoine et al 2009; Barde-Cabusson et al  2012; Lénat et al 2012a).

 Therefore, in this study, we consider that the main SP generation is due to the strongest hydrothermal fluids flowing through the permeable structures within the hydrothermal system, independently of the hydrothermal system boundaries.

Classical self-potential studies can only determine the location of the surface projection of the underground water structure. In order to provide quantitative information regarding the depth of the main water flow SP data can be processed by multi-scale wavelet tomography using two to four series of analyzing wavelets to define a domain of validity of depth for each of these hydrothermal activity sections. The vertical movement of the potential generation area (SP source) associated with the hydrothermal activity sections can then be evaluated in order to investigate the temporal behavior of the hydrothermal system.

### 4.2. Multi-scale wavelet tomography

Multi-scale wavelet tomography (MWT) is a signal processing method based on continuous wavelet theory (e.g. Grossmann & Morlet 1984; Saracco 1987; Saracco et al 1989) and potential theory (Poisson operator; e.g. Courant & Hilbert 1962; Saracco et al 1990; Moreau et al 1997, 1999). The general mathematical expression of the continuous wavelet transforms applied to potential fields is presented in the work of  Moreau et al. (1997) and can be found in Appendix B. The wavelets based on the Poisson kernel family are specific for analyzing potential field data because they regroup properties from both wavelet and  potential  field  theories (Saracco & Tchamitchian 1990; Moreau et al 1997). In the literature, other examples of wavelet analyses can found on potential field data (Hornby et al 1999; Martelet et al 2001; Fedi et al 2004, 2010). Synthetic examples for MWT method using four analyzing wavelets derived from the Poisson kernel family coupled to a statistical approach, to define a domain of validity of depth estimate of  potential sources are in Appendix B.

In this study, we used four real wavelets based on the Poisson kernel family (Saracco & Tchamitchian 1990; Moreau et al 1997; Saracco et al 2004, 2007; Mauri et al 2010, 2011, 2012; Caudron et al 2016) which are the second and third vertical derivative (V2 and V3, respectively) and second (γ = 2) and third (γ = 3) horizontal derivative (H2 and H3, respectively), where γ is the order of the derivative. In order to efficiently apply a continuous wavelet transform, the analysis is done within the frequency domain, rather than in the spatial domain (see Appendix B). Several hundred dilations are used to detect and characterize all the singularities shaping the analyzed SP signal (e.g. Moreau et al. 1997; Saracco et al. 2004; Mauri et al. 2010).

The output of the multi-scale wavelet tomography over a full range of dilations generates a map of wavelet coefficients in the space (x,z) (Fig.3), where minima and maxima of the correlation coefficient are organized on lines of extrema. These lines are organized in pairs or triplets, which converge in a cone shape structure toward the electrical generation area (Fig.3). Each wavelet acts as a specific filter that responds slightly differently to the main electrical generation area, however the choice of the analyzing wavelet is not trivial.  When using only one real analyzing wavelet deriving from Poisson kernel, results can be affected by a significant uncertainty in the location of the main electrical generation area due to the secondary source effect (Mauri et al 2010, 2011). To decrease this uncertainty, the MWT method uses a statistical approach by calculating the depth and its uncertainty using results from several  analyzing wavelets (Mauri et al 2010, 2011). Doing so,  the main electrical  generation area is  better constrained by 1) reducing the uncertainty on the calculated depth and 2) reducing the effect of secondary electrical generation area. The statistical approach allows for definition of a volume in space, which contains the main electrical generation area, through the constraint of the standard deviation value of the depth.

In our study, the measured SP signal corresponds to the electrical field, which is the sum of different electrical generation contributions, therefore, the wavelet analyzes locate at depth the strongest electrical generation area (e.g. electrokinetic effect associated to hydrothermal fluid movement, thermo-electric effect, resistivity contrast). Secondary effects such as local infiltrating meteoric water act as noise in the depth estimation  of the main hydrothermal activity section (e.g. results of 1998 SP analyzes; Fig. 2).  Furthermore, past studies have shown that: (1) in the case of PdFV, the electrokinetic effect is the main component of electrical potential generation (e.g. Michel and Zlotnicki 1998;  Barde-Cabusson et al 2012) and (2) that the electrical potential from resistivity contrasts has a negligible effect on MWT depth determination (Mauri et al 2010,  2011). For more information, examples are given in Appendix B. Therefore, the MWT method gives information on where the electrokinetic effect associated to hydrothermal fluid movement is the strongest (Mauri et al 2010, 2011). The calculated depth value and its associated uncertainty value can then be interpreted in terms of the section along the hydrothermal flow path where hydrothermal circulation is the strongest.

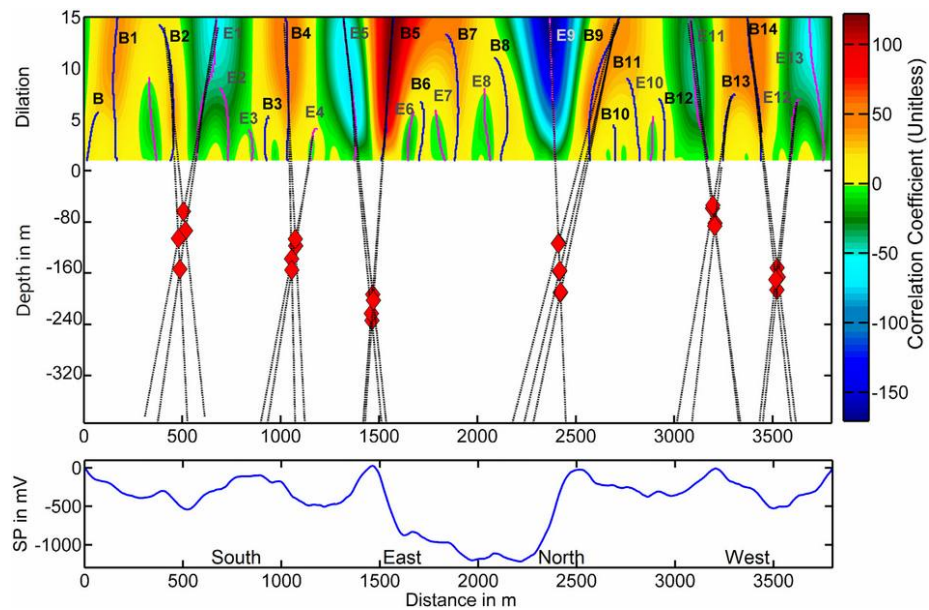

Fig. 3. Example of depths (red diamonds) calculated from multi-scale wavelet tomography (MWT) on the November 2003 SP profile (bottom) of Piton de la Fournaise volcano. Analysis was made with the V2 wavelet along 500 dilations over a dilation range from 1 to 15. B to E are lines of extrema.

5. Self-potential and MWT results

The self-potential data set used for this study consists of 10 complete profiles surveyed around the summit craters of Dolomieu, Bory over 16 years (1993 to 2008) (Fig2). The profiles were acquired in June 1993, December 1995, March 1998, December 1999, April 2010, April 2003, November 2003 (Malengreau et al 1994; Lénat et al 2000; Labazuy, internal reports), March 2005 (this study) July 2006 and May 2008 (Barde-Cabusson et al 2012). The sampling interval of these surveys was 12.5 m or 25 m along the summit loop (~3.7 km in lenght). All SP profiles are relative to the Bory Reference Station (BRS) on the south rim of the Bory crater (Fig. 1c). SP profiles, prior to 1993 and that of 2001, are not considered here as they had excessively large sampling steps or did not completely encircle the summit craters. Over the 16 years of self-potential surveys, 5 of the 10 surveys March 1998, December 1999, April 2002, 2003 and November 2003) were made less than a few weeks before, during or after an eruption at the summit cone. During the March 2005 survey, an eruption was ongoing in the north-east part of the Enclos Fouqué caldera (Fig.1b). 2006 and 2008 are particularly interesting as these SP surveys pre- and post-date, respectively, the collapse of Dolomieu crater, which occurred during the April 2007 eruption.

Over the years, each SP survey had a drift of 100 mV at the end of the 3.7km loop profile. Drift of the electrodes was constrained twice a day. Following drift correction, the uncertainty on the collected data is estimated to be 10 mV. In 2005, the closure error was b50 mV. The traditional leapfrog technique (e.g. Corwin & Hoover, 1979) was used for each of SP surveys in this work. In 2006 and 2008, the same method was used and a complete description can be found in the work of Barde-Cabusson et al 2012. Error and noise on data acquisition, over the 16-year period, are negligible in comparison to the amplitude of the SP signal.

Several persistent anomalies can be identified in the SP profiles (Fig.2) and are associated to structural limits (collapsed high permeability structure, pit craters, faults) (Malengreau et al 1994; Barde-Cabusson et al 2012). In agreement with the notation used in Barde-Cabusson et al 2012, the 6 main positive SP anomalies are labelled as follows: the Bory north anomaly, BN; Bory south anomaly, BS; Dolomieu south anomaly, DS; Dolomieu south-east anomaly, DSE; Dolomieu east anomaly, DE; and the Dolomieu north anomaly, DN. In addition, one large negative anomaly (EAST) is located on the eastern part of the summit crater (Fig. 2). EAST has a negative SP/elevation gradient and thus represents down flowing water (Saracco et al 2004; Barde-Cabusson et al 2012).

Although the BRS reference station is located on the summit cone and as such, its absolute SP value may vary over time, any change occuring at the reference station will not affect the relative amplitude of the SP anomalies or the shape of the signal. Furthermore, Multi-Scale Wavelet Tomography of self-potential profiles is effective with relative SP data (Mauri et al 2010, 2011, 2012; Caudron et al 2016), which considers peak to peak amplitude rather than absolute amplitude. Consequently, the MWT-derived depths for the 10 SP profiles are independent of the location of the chosen reference point.

In order to characterize the effect of noise in each SP profile on the MWT depth calculation, the signal/noise ratio (SNR) of each of the 10 profiles was calculated. In a related study, Mauri et al 2010 show that a synthetic SP signal with SNR between 10% and 20% of Gaussian white noise will cause errors in the MWT-calculated depths of 25 m for electrical generation areas located at 200 m depth. If the measured potential field is contaminated by high levels of noise (e.g. due to poor sampling, unstable measurements), a filtering method based on orthogonal wavelets (as Moreau et al 1996) can be applied prior to MWT processing. In this study, the SNR for each SP profile was determined by calculating the ratio using as reference the smoothed SP signal, which has been filtered to remove high frequency variations. All the 10 SP profiles have between 1% and 16% noise, which allows us To be confident that noise does not have a significant influence on the calculated depth.

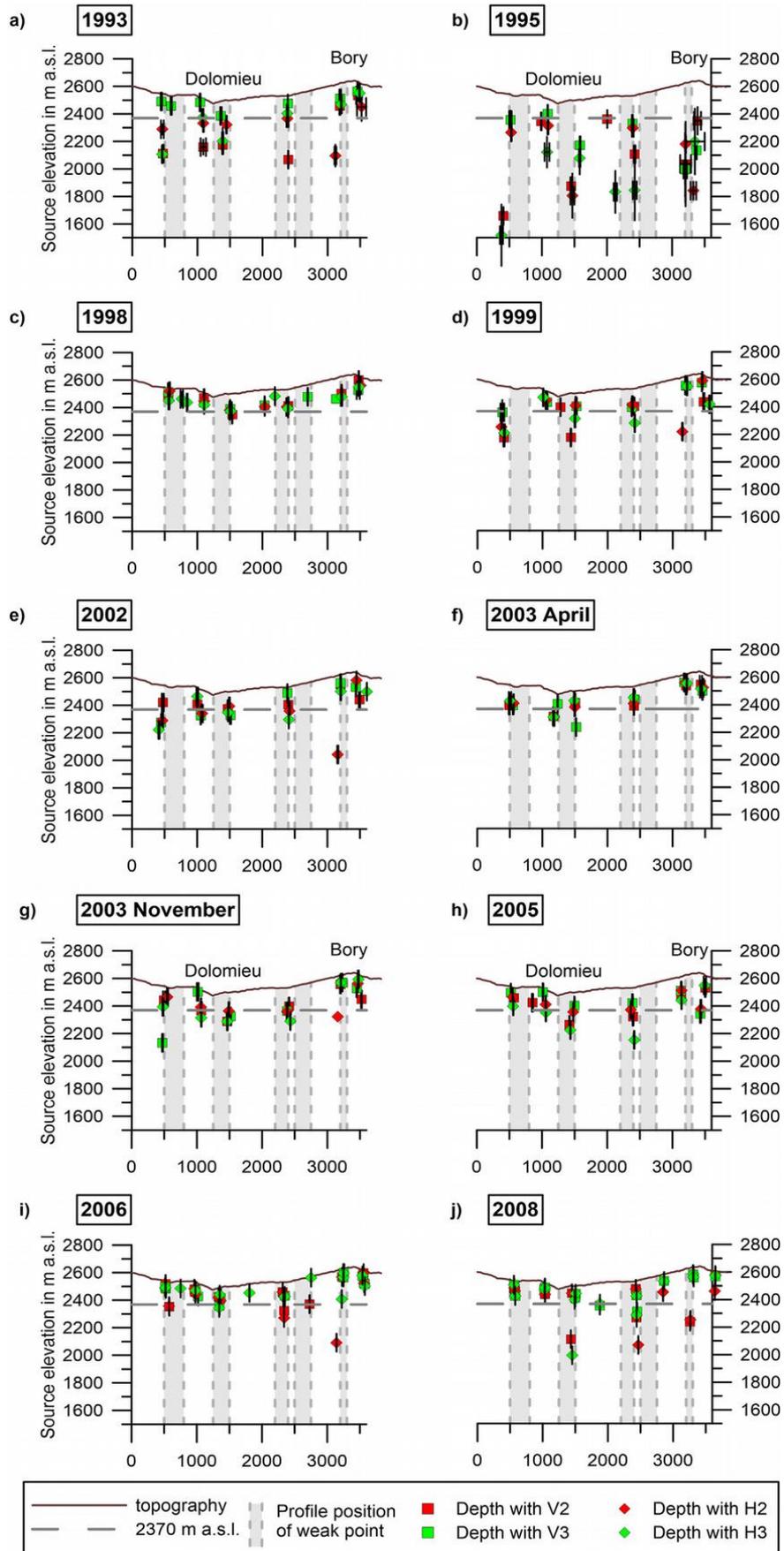

Fig. 4. Multi-scale wavelet tomography-calculated depths for the 1993 to 2008 SP profiles. V2, V3 are the second and third vertical derivative wavelet, respectively. H2 and H3 are the 2nd and 3rd horizontal wavelet, respectively. For each year, there are 24 depths found along the profile. Error bars represent the uncertainty on the best fit intersection see Fig. 3). The horizontal dashed grey line is the global average depth based on 238 values found over the 16 years (Table 1) and which represent median elevation of the summit cone.

Each wavelet analyzis was calculated for each profile, with 500 dilations on a range from 1 to 15. Only depths consistently inferred by at least three of the four wavelet analyzes are considered significant in this study. The depths obtained from all the analyzes are organized in specific areas around the crater, which we call the electrical generation area. In total, 6 main electrical generation areas were located with each the four wavelets on each of the 10 SP profiles for a total of 238 MWT calculated depths (Fig. 4a,b; Table 1). The uncertainty of depths (di-amonds and squares, Fig. 4) is due to multiple best fit possibilities and is represented with error bars.

In order to better evaluate temporal variations, the depth values for each of the 6 electrical generation areas are also presented as an average depth value (with its associated standard deviation Table 2, Fig.5). Furthermore, an arbitrary average depth boundary equal to the average of 238 depths is defined to be at 2380m a.s.l. or 174m below the topographic surface (Fig5). This average depth also corresponds to the median elevation of the summit cone, which rises around 2600m a.s.l. (Michon et al 2009a,b). Interestingly, this average from ~2200m a.s.l. to depth is very similar to that of a hydrothermally-altered layer and fumarole belt found around the crater wall following the 2007collapse of Dolomieu crater (Staudacher 2010; Peltier et al 2012). The hydrothermally-altered layer was reported to represent the effect of both hydrothermal vapour and high temperatures (100–200 °C). For this study, we define all depths above this boundary (~2400m a.s.l.) as shallow and those below it, as deep.

For the 1993 SP profile, depth values are spread equally on either side of  this average depth boundary,  from 2100 m to the surface (Fig. 4a). In1995, depth values are more scattered, but consistently below the average depth boundary, ranging from ~1500 m to ~2400 m a.s.l. (Fig.4b, Table 2). In 1998, the majority of depth values are above the average depth boundary,  from a few tens of meters below the surface to 2150 m a.s.l. (Fig. 4c). In 1999, as in 1993, depth values scatter on both sides of  the average depth boundary, from ~2600 m down to ~2200 m (Fig. 4d). From April 2002 to 2008, depth values are spread around the average depth boundary from ~2150 m to near the surface (Fig. 4e,f and g–j).

## 6. Discussion

### 6.1. Hydrothermal activity sections

Previous studies demonstrated the existence of one large hydrothermal system characterized at the surface by six main and persistent SP anomalies (Malengreau et al 1994; Lénat et al 2000, 2012a; Saracco et al 2004; Barde-Cabusson et al 2012). This interpretation was confirmed with the 2006 survey which demonstrated that these six SP anomalies were connected to the SP anomalies extending throughout the craters (Barde-Cabusson et al., 2012).

Each of these six main SP anomalies characterize a part of the hydrothermal system, which can then be considered in terms of a hydrothermal  zone, where each zone is associated to an area of structural weakness, expressed at the surface by faults and pitcraters (Fig.1 c,d) and the hydrothermally-altered zone within the crater wall. Therefore, the six SP anomalies indicate preferential pathways ways for both magmatic and hydrothermal fluids (Lénat et al 2000 2000, 2012a; Saracco et al 2004; Barde-Cabusson et al 2012; Peltier et al 2012). Over time, each of these hydrothermal zones is defined by its own characteristic SP anomaly, which is well defined in both amplitude and space on the summit cone (Fig. 1c).

Based on our results, the determined depths over the 10 SP profiles can also be grouped into 6 electrical generation areas representing the areas of strongest electrokinetic effect processes within the summit cone. Therefore along the crater rims, these 6 electrical genera-tion areas are interpreted to be where the strongest hydrothermal circulation is taking place within the summit cone. Since each of the six electrical generation areas matches the location of existing hydrothermal zones (Fig.5), it is reasonable to consider that these electrical generation areas are 1) expressions of hydrothermal circulation at depth, and 2) locate the section along the hydrothermal path where the flow is the strongest. Consequently, each of the electrical generation areas located by our MWT analyzes is named according to its associated hydrothermal zone (Fig. 5); BN is the Bory north anomaly; BS is Bory south anomaly; DS is Dolomieu south anomaly; DSE is Dolomieu south-east anomaly; DE is Dolomieu east anomaly, and DN is the Dolomieu north anomaly).

In order to evaluate the temporal change of each of these zones, the average depth of each electrical generation area for each year is presented Fig.5. In these depth estimates, the uncertainty is generally small, however, the scatter can be large (ranging from 18 m to 270 m; Fig.5, Table 2). In order to identify similar temporal behaviors of the hydrothermal activity sections, a coefficient of determination (r) was calculated by comparing the time-series depths of the Bory South (BS) hydrothermal activity section, taken as a reference,  with each time-series depth of the other 5 electrical generation areas (DS, DSE, DE, DN, BN). Although the choice of Bory South as reference is arbitrary, the coefficient of determination (r) between BS and each of the other potential generation areas (DS: 0.83, DSE: 0.77, DE and DN: 0.98, BN: 0.94), suggests that all hydrothermal activity zones have a common dynamic behavior (Table 2, Fig5). Similar hypotheses have been suggested for the hydrothermal variation inferred from SP data for the period between 2006 and 2008 (Barde-Cabusson et al., 2012).

### 6.2. Effect of rainfall

Historical rainfall records across the Island of Réunion show that heavy rainfall events can be several days in duration (Barcelo et al., 1997). The eastern side of the island is more affected by precipitation than the other parts of the island (Barcelo and Coudra, 1996); the rainfall gradient reaches its maximum between 1300 and 1800 m a.s.l. within the Enclos Fouqué caldera, between these elevations, the maximum rainfall reaches $12m \dot{y}r$ . Below 1300m and above 1800m a.s.l., the rainfall decreases to $~7m yr$ (Barcelo et al 1997). As suggested by Malengreau et al. 1994, in the summit area above 2300 m a.s.l., the amount of rainfall would be the same along the entire SP loop. Consequently, all the electrode and reference points would be affected by similar variations in electrical potential. Between 1993 and 2008, except during the 1998 survey, the SP surveys (Fig. 2) show no significant noise. In the case of the 1998 SP profile, it is difficult to determine exactly the nature of the strong noise. However, an analysis of this noise shows that it is stronger on the eastern side of the summit, and thus likely due to water infiltration associated to a strong rainfall event. Based on our results in terms of depth estimation, the increase noise on the SP profile is expressed through an increase in scattering of the depth values (Fig. 4a, b).

### 6.3.Relationship between inferred hydrothermal  depths and  Volcanic activity

From its base (at sea level) to its highest point (2600 m a.s.l.), the summit cone of Piton de la Fournaise volcano is cut by two main structures: 1) a collapsed high permeability structure with ring faults and 2) proximal dyke pathways, which feed each of the summit eruptions and host the main hydrothermal reservoir (Peltier et al 2009, 2012; Lénat et al 2012a). The top of the main hydrothermal reservoir is estimated to be around 1000 m below the surface, i.e., from 1300 to 1500m a.s.l. (Lénat et al 2012a and references therein). In addition, following The 2007- collapse event, evidence of shallow active hydrothermal circulation is found at 2340 m a.s.l., such as persistent fumaroles

Table 1

Depths of the hydrothermal zones calculated by multi-scale wavelet tomography of self-potential profiles on Piton de la Fournaise volcano between 1993 and 2008. Number of wavelets used in MWT to locate the potential generation area along profile. σ is one standard deviation.

| Hydrothermal zone | Date | Number | Distance in m | σ x in m | Depth Z in m | σ Z in m | Elevation in m a.s.l |
|---|---|---|---|---|---|---|---|
| DS | June 1993 | 4 | 461 | 16 | −275 | 163 | 2272 |
| | November 1995 | 4 | 471 | 69 | −524 | 391 | 2022 |
| | March 1998 | 4 | 564 | 9 | −54 | 36 | 2477 |
| | November 1999 | 4 | 393 | 22 | −301 | 90 | 2250 |
| | April 2002 | 4 | 451 | 26 | −247 | 81 | 2301 |
| | April 2003 | 4 | 526 | 32 | −132 | 35 | 2408 |
| | November 2003 | 4 | 500 | 32 | −184 | 144 | 2362 |
| | March 2005 | 4 | 550 | 19 | −79 | 36 | 2454 |
| | July 2006 | 4 | 537 | 26 | −67 | 63 | 2455 |
| | May 2008 | 4 | 569 | 8 | −48 | 29 | 2470 |
| DSE | June 1993 | 4 | 1075 | 27 | −162 | 116 | 2350 |
| | November 1995 | 4 | 1071 | 42 | −259 | 141 | 2254 |
| | March 1998 | 4 | 1104 | 4 | −81 | 21 | 2427 |
| | November 1999 | 3 | 1078 | 88 | −69 | 18 | 2443 |
| | April 2002 | 4 | 1034 | 34 | −137 | 56 | 2380 |
| | April 2003 | 4 | 1190 | 27 | −151 | 51 | 2339 |
| | November 2003 | 4 | 1058 | 20 | −132 | 61 | 2382 |
| | March 2005 | 4 | 992 | 88 | −96 | 53 | 2433 |
| | July 2006 | 4 | 980 | 17 | −46 | 16 | 2462 |
| | May 2008 | 4 | 1034 | 3 | −32 | 21 | 2466 |
| DE | June 1993 | 4 | 1395 | 33 | −224 | 105 | 2271 |
| | November 1995 | 4 | 1537 | 69 | −495 | 200 | 2007 |
| | March 1998 | 4 | 1516 | 17 | −142 | 45 | 2358 |
| | November 1999 | 4 | 1491 | 35 | −173 | 99 | 2325 |
| | April 2002 | 4 | 1487 | 17 | −137 | 47 | 2361 |
| | April 2003 | 4 | 1503 | 11 | −13 | 81 | 2366 |
| | November 2003 | 4 | 1485 | 23 | −184 | 39 | 2314 |
| | March 2005 | 4 | 1458 | 31 | −187 | 75 | 2315 |
| | July 2006 | 4 | 1349 | 13 | −72 | 30 | 2403 |
| | May 2008 | 4 | 1473 | 27 | −131 | 168 | 2337 |
| DN | June 1993 | 4 | 2387 | 9 | −202 | 164 | 2329 |
| | November 1995 | 4 | 2414 | 27 | −488 | 265 | 2045 |
| | March 1998 | 4 | 2285 | 166 | −127 | 19 | 2459 |
| | November 1999 | 4 | 2393 | 20 | −157 | 61 | 2374 |
| | April 2002 | 4 | 2403 | 13 | −146 | 74 | 2386 |
| | April 2003 | 4 | 2408 | 7 | −110 | 27 | 2422 |
| | November 2003 | 4 | 2404 | 22 | −180 | 45 | 2352 |
| | March 2005 | 4 | 2388 | 19 | −214 | 106 | 2318 |
| | July 2006 | 4 | 2335 | 13 | −118 | 68 | 2400 |
| | May 2008 | 4 | 2441 | 12 | −174 | 131 | 2344 |
| BN | June 1993 | 4 | 3180 | 34 | −217 | 163 | 2404 |
| | November 1995 | 4 | 3247 | 112 | −524 | 242 | 2102 |
| | March 1998 | 3 | 3207 | 28 | −138 | 28 | 2488 |
| | November 1999 | 4 | 3197 | 27 | −137 | 134 | 2488 |
| | April 2002 | 4 | 3192 | 20 | −210 | 221 | 2414 |
| | April 2003 | 4 | 3199 | 17 | −70 | 17 | 2556 |
| | November 2003 | 4 | 3204 | 19 | −83 | 70 | 2543 |
| | March 2005 | 4 | 3136 | 5 | −126 | 33 | 2487 |
| | July 2006 | 4 | 3239 | 12 | −61 | 60 | 2553 |
| | May 2008 | 4 | 3294 | 23 | −145 | 154 | 2464 |
| BS | June 1993 | 4 | 3383 | 159 | −189 | 198 | 2449 |
| | November 1995 | 4 | 3378 | 59 | −499 | 251 | 2138 |
| | March 1998 | 4 | 3484 | 19 | −71 | 38 | 2556 |
| | November 1999 | 4 | 3485 | 49 | −122 | 75 | 2507 |
| | April 2002 | 4 | 3481 | 58 | −112 | 51 | 2518 |
| | April 2003 | 4 | 3450 | 19 | −113 | 26 | 2522 |
| | November 2003 | 4 | 3478 | 28 | −97 | 50 | 2536 |
| | March 2005 | 4 | 3466 | 36 | −180 | 106 | 2454 |
| | July 2006 | 4 | 3559 | 13 | −57 | 35 | 2551 |
| | May 2008 | 4 | 3645 | 3 | −64 | 46 | 2551 |

and thermal anomalies (Staudacher 2010; Peltier et al 2012). These observations correlate well with our MWT depth results.

Based on our results, the MWT depths and the coefficient of determination ($\hat{r}^2$) of the 6 potential generation areas, three periods can be characterized. In June 1993, approximately 10 months after the last eruption prior to the period of quiescence, the potential generation areas, inferred to represent the main hydrothermal activity sections, are confined near the median elevation of the summit cone, where later in 2007 post-collapse event, the fumarole belt is located (~2400m a.s.l., Table2, Fig.5). By 1995, 26 months after the August 1992 eruption, the hydrothermal activity sections had dropped to

~2100m a.s.l. suggesting a significant change within the hydrothermal system. With such a long period without new magmatic intrusion into the summit cone, it is reasonable to suggest that the hydrothermal system lacked sufficient energy to continue rising toward the summit. This hypothesis would explain the increase in the inferred depth. In terms of process, it is not unreasonable to assume that the hydrothermal system was cooling with depressurization of its fluids during the period of 1995 (Table1, Fig.5). With the resurgence of eruptive activity in March 1998 (Peltier et al 2009), frequent eruptions took place on the summit cone (above 2200m a.s.l., red bars on Fig.5). Magma migration along dykes through the summit cone would have increased the heat flux through

Table 2
Minima, maxima, mean and σ of the depths per year of the 6 hydrothermal zones on Piton de la Fournaise volcano between 1993 and 2008. σ is one standard deviation.

| Date | Depth in m a.s.l. | | | | Number of depth values |
|---|---|---|---|---|---|
| | Min | Max | Mean | σ | |
| June 1993 | 2068 | 2563 | 2360 | 155 | 24 |
| November 1995 | 1519 | 2401 | 2097 | 244 | 24 |
| March 1998 | 2156 | 2600 | 2448 | 84 | 23 |
| November 1999 | 2177 | 2590 | 2394 | 125 | 23 |
| April 2002 | 2043 | 2581 | 2393 | 123 | 24 |
| April 2003 | 2237 | 2562 | 2434 | 89 | 24 |
| November 2003 | 2132 | 2593 | 2410 | 116 | 24 |
| March 2005 | 2154 | 2550 | 2408 | 99 | 24 |
| July 2006 | 2400 | 2553 | 2466 | 64 | 24 |
| May 2008 | 2337 | 2551 | 2438 | 83 | 24 |

the upper part of the summit cone (*Saracco et al 2004; Lénat et al 2012a*). Based on our results, this change in activity coincided with the upward movement of hydrothermal fluids, by almost 400m, to ~2400m a.s.l. (maxima at 2600m a.s.l. Table 2, Fig.4a–d). From 1999

1999 to 2008 (end of this study) the eruptive activity was high (with 34 eruptions occurring above 2200 m a.s.l. since 1998; Peltier et al 2009) implying a potentially high heat flux inside the summit cone. During the same time, our results suggest that the depth of the hydrothermal activity section remained near or above the 2400 m a.s.l. limit (Table 2, Fig.5). Consequently, for the 1993–2008 period, our results show that the inferred MWT depths give qualitative information on changes within the hydrothermal system and that the depth of the hydrothermal system is directly affected and controlled by the volcanic activity within the summit cone of Piton de la Fournaise volcano.

### 6.4. MWT-inferred depths and its use in investigating changes in hydrothermal systems

Several conditions are inferred to explain the large scattering of the MWT depths during the 1993–1998 period (Fig.4):

1) One of the main factors considered is the complex geology of the collapsed structure, which consists of brecciated layers of dense lava flows interbedded with more permeable scoria layers. The collapsed structure is further cut by dense basaltic intrusions forming dykes and sills. Finally, over time, hydrothermal fluids will alter and change the physical properties of the different units. This hydrothermal alteration will not be homogeneous due to the large permeability contrast between the different rock units as well as the fact that the summit cone is cut by numerous faults and fractures following the structural limits. This leads to a very complex network of hydrothermal circulation pathways within the summit cone (Lénat et al 2012a; Peltier et al 2012).

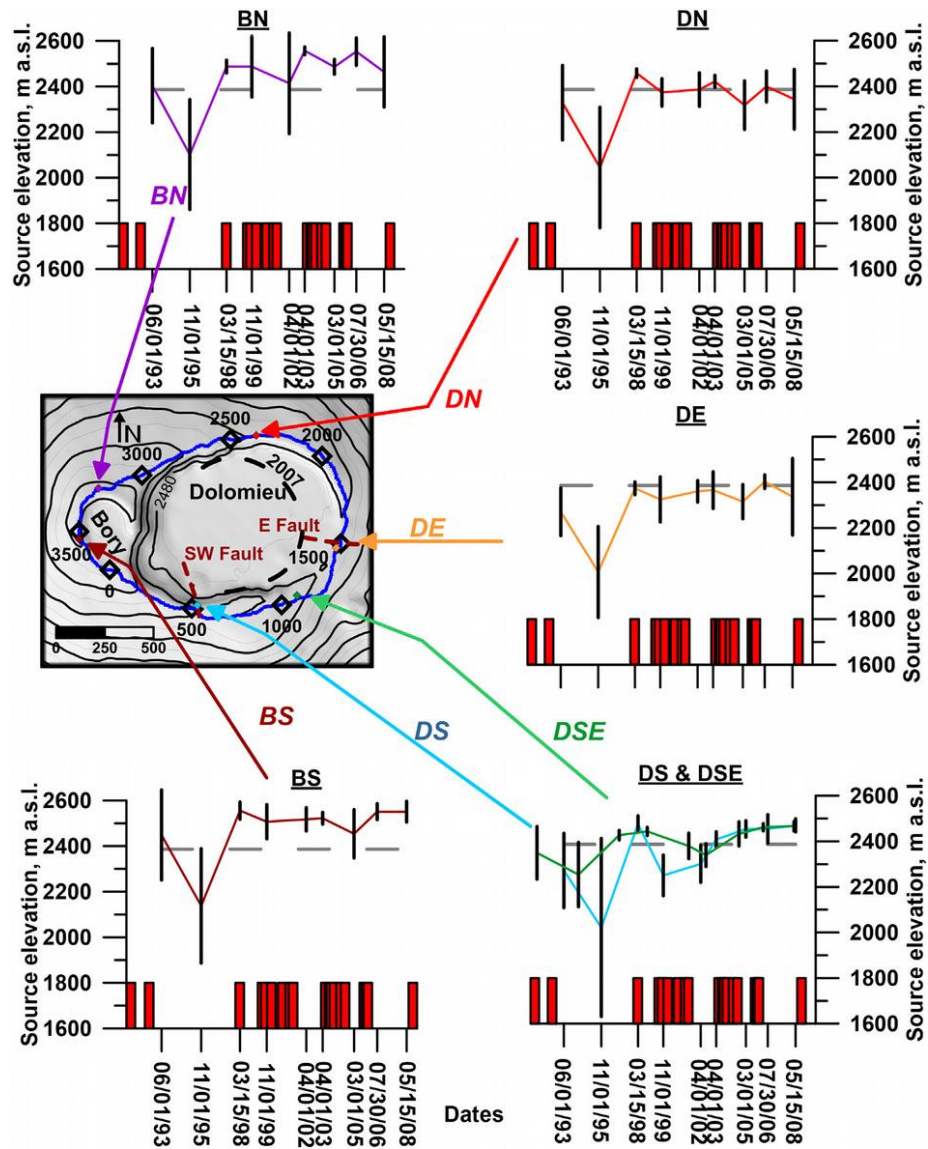

Fig.5. Locations of the main hydrothermal structures found by SP and MWT on SP profiles along the survey loop around the summit Bory-Dolomieu craters and their time-series depth evolution over the period from 1993 to 2008. Dashed grey line represents both the average global depth over the 16 years and the median elevation of the summit cone. Red bars denote eruptive events that took place above 2200 m a.s.l. on the summit cone. The black dashed line represents limit of the 2007 collapse event.

2) The nature of the hydrothermal fluids may distort the SP signal. The potential generation area of the analyzed SP signal consists of a volume of fluid extending along structural limits (i.e. fractures, faults or permeable layers). The wavelet analyzes allow us to estimate the depth of electrical generation area as a localized punctual point rather than a volume. However, using four analyzing wavelets (Fig.4) we statistically calculated the uncertainty in both x and z- axes and then interpret in terms of extent of where the hydrothermal circulation is the strongest (Fig.5). Therefore, the information carried by 4 series of analyzing wavelets of different orders define a domain of validity for the estimated depths for each hydrothermal activity section.

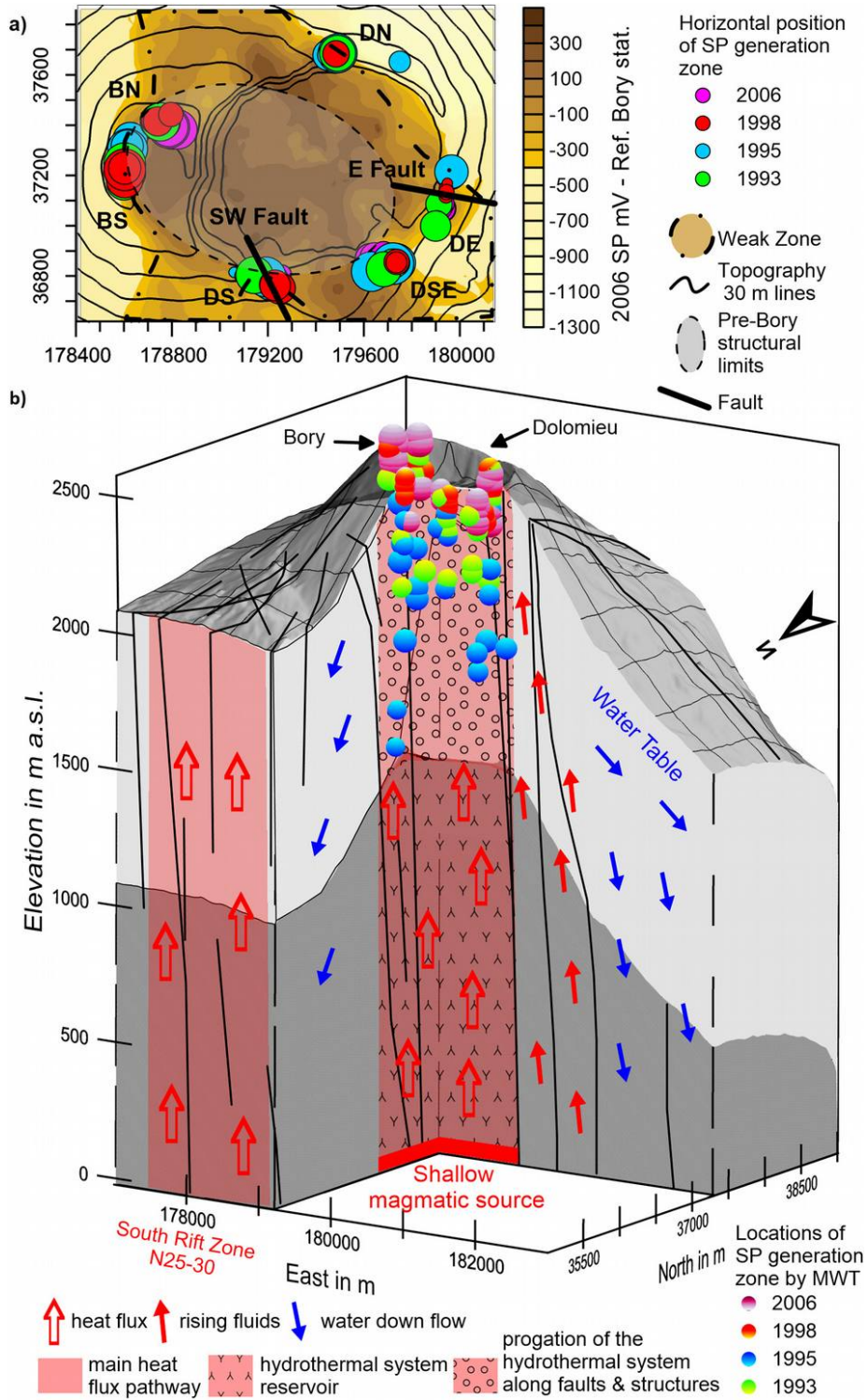

Fig. 6. 3D schematic view of the hydrogeological structures within the summit cone of Piton de la Fournaise volcano, as proposed by the works of Lénat et al, 2012a and Barde-Cabusson et al 2012, and the hydrothermal activity sections determined by this study for 1993, 1995, 1998 and 2006. a) Locations of the 2006 SP anomalies, faults and structures with the horizontal locations of the SP generation zones. b) 3D view of the locations at depth of the SP generation zones, which are calculated by MWT. The SP generation zones are where the electrical generation occurs within the hydrothermal system, and how SP generation zones correlate with the known hydrogeological structures. See Table 1 for details.

## 6.5. Depths of electrical potential generation area and 2007 collapse

In order to provide a clear explanation of the depth value of the electrical potential generation areas obtained through MWT in 2008,  it is first necessary to understand the volcanic activity in 2007 and 2008. In April 2007, following intense eruptive activity from vents on the lower flank of the Grand Brûlé, the inside of

Dolomieu crater collapsed into the magma chamber (Staudacher et al 2009). The collapsed rock column left a crater 340m deep. It is significant to note that the products of this event consisted of only phreatic ash (Staudacher et al 2009) and the collapsed area did not affect the entire Dolomieu crater (Fig. 5), dashed line in the crater map) leaving an eastern plateau (Barde-Cabusson et al., 2012).

2008 was marked by two main periods. From April to July 2008, the summit was subsiding with a reduction of crater size (from 1 to 3 m following the 2007 collapse) and a subsidence (from 40 to 99 cm since the 2007 collapse) (Staudacher, 2010). During this period, the background seismicity was normal with b10 volcano-tectonic seismic events per day (OVPF internal report, 2008). In August and September 2008, seismic activity increased, indicating magma transfer within the volcanic edifice. Finally, on September 21, 2008, following an important increase in seismic activity, an effusive eruption occurred within Dolomieu crater (Staudacher, 2010).

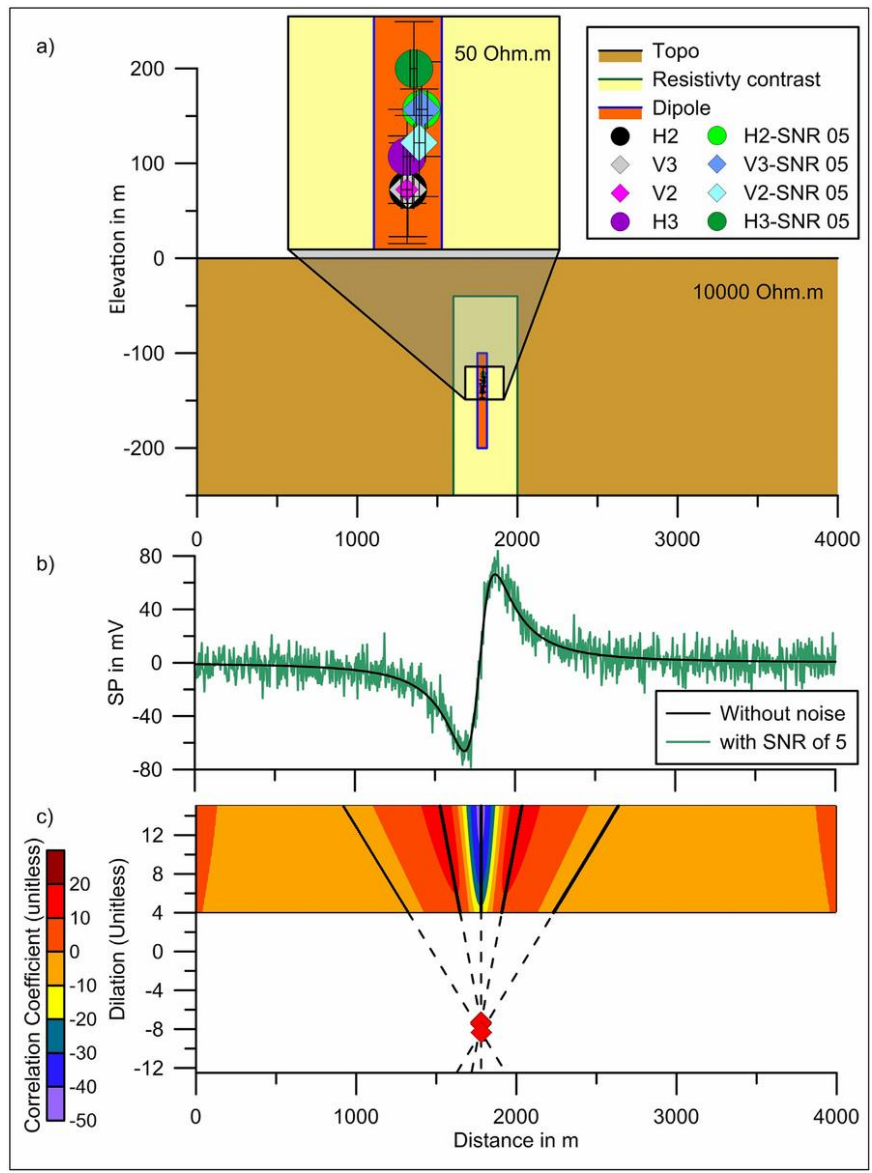

Fig.7. Synthetic signal analyzed with MWT method. a) 2D model to generate SP signal: a dipole (red rectangle 100 m × 60 m) is embedded within a low resistivity layer (50 Ωm) and surrounded by a high resistivity layer (10,000 Ωm). Topography is flat and the contrast resistivity contour (green line) represents the interface between the two mediums of differing resistivity. The top of the dipole is at 100m below the topographic surface. The depth values are the average calculated depth obtained for each of the 4 wavelets (H2, H3, V2, V3) for both signal with and without noise. b) Synthetic self-potential signals generated by a dipole with noise (SNR 5) and without noise. c) Example of calculated depths of vertical dipole (as shown in a), when depth is calculated with the wavelet H3 on the real part of the continuous wavelet analyzis. The analysis has been done for 500 dilations over a range of dilations from 4 to 15. The black lines are the extrema lines, the dashed lines are the linear regressions of the extrema and the red diamonds are the calculated depths.

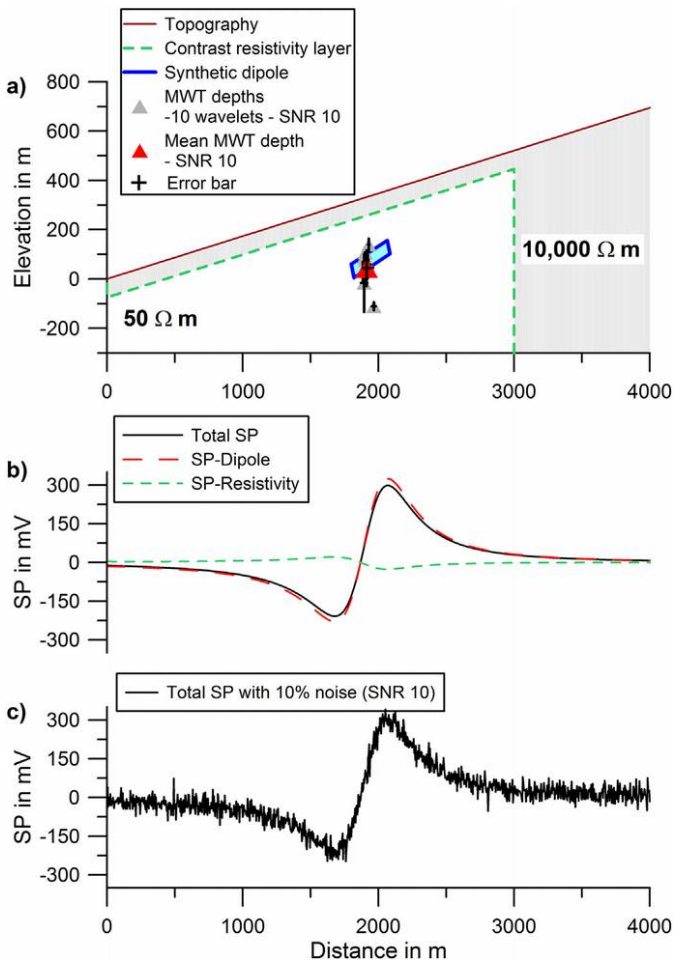

Fig.8. The effect of 10% noise (SNR 10) on a synthetic SP signal generated by a dipole in a non-uniform medium and in a 2D model. Signal is calculated with a sampling rate of 1m. *a)* 2D model used to generate self-potential signal has a slope of 10° and two media of different resistivity; the dashed line represents the limit between the two media. The orientation of the dipole is horizontal (blue rectangle, 10 m × 60 m). The dipole depth is defined by its upper left corner (390 m below the topographic surface) and its bottom left corner (400 m below the topographic surface). *b)* Synthetic self-potential signal generated from each component of the model. The signal generated by the dipole is the blue dashed line, while the green dashed line is the effect of the contrast resistivity layer. The total SP signal is presented by the solid black line. c) 10% Gaussian noise (SNR = 10) applied to the total SP signal presented in b and used to compute the depths shown in a). Modified after Mauri et al. (2011).

The May 2008 SP survey conducted by *Barde-Cabusson et al 2012*, took place during the period of deflation. From a stress field point of view, the deflation of the summit is likely to have an impact on the fractures and fluid circulations at depth. A comparison between SP surveys of July-

August 2006 and May 2008 shows that the main SP variations are not located along the main fault structures (DS, BN, DN, DE) (Barde-Cabusson et al 2012). Indeed, near the collapsed area, DS and BN show very little decrease of SP (≤100 mV), when separated from the collapse by the eastern plateau, DN and DE present a slight SP increase (≥200 mV).

Our MWT results for 2006 and 2008, show that the depths of potential generation areas associated to DS, BN, DN, DE fault structures are similar between the 2 surveys (Fig .5), which correlates well with the expected small SP variation. Based on 1) the subsidence of the summit, 2) the relatively low SP variations along the main faults (DS, BN, DN, DE) and 3) the relative stability of calculated depth of the hydrothermal activity section, it is reasonable to consider that the collapse did not significantly decrease the hydrothermal activity along the main faults and that the deformation occurring since 2007 contributed to maintain the hydrothermal system under pressure.

### 6.6. MWT-inferred depths as a complementary tool for volcano monitoring

Previous work on the 2007 collapse of the summit crater has shown that the hydrothermal system played a significant role during this collapse event, which marked the end of the 9-year eruptive cycles (*Barde-Cabusson et al 2012; Lénat et al 2012a*). Indeed, expansion of hydrothermal activity will weaken the volcanic edifice by generating hydraulic fracturing, or by boiling-induced decompression within the shallow magma reservoir (*Lénat et al 2012a*). Consequently, hydrothermal activity facilitates intrusion when magma rises from the shallow magma reservoir. During the period of quiescence of Piton de la Fournaise (1993–1998), our results suggest that the electrical generation areas were deeper (from 200 m to 1000 m below the surface, Fig.5). In 1995, among the electrical generation areas associated with Dolomieu crater, the DS, DE & DN electrical generation areas appear to be the deepest (Fig. 5, Table 2). From the structural information, these electrical generation areas are located near the crater's ring faults, which reach the hydrothermal reservoir at depth (Fig.5). With the resumption of eruptive activity in 1998 the inferred depths representing the hydrothermal activity appears to be shallower and reaches a similar elevation range between 2300 and 2500m a.s.l. (Table2).

Between 1993 and 1998, rising from the main hydrothermal reservoir, smaller hydrothermal activity sections were spread at different levels along the weak zones corresponding to structural limits of the collapsed structure (Fig6). A realistic explanation of the increased depth of the hydrothermal activity sections is that lower magmatic activity within the summit cone resulted in lower energy supply to the hydrothermal system. The temperature and pressure of the hydrothermal fluids thus could not sustain the fluids in the upper part of the summit cone (Fig.6). With the resumption of eruptions in the summit area after 2000, it is likely that the heat flux reaching the entire hydrothermal system significantly increased. This period corresponds to the pre-eruptive or crisis period described in Saracco et al (2004) and Lénat et al. (2012a). Our results show that since 2000, all the MWT depths formed small clusters located near or above the 2400 m a.s.l. mean reference level.

Multi-scale wavelet tomography of SP signals is a robust tool to investigate and monitor internal changes that may occur within hydrothermal systems and brings new insights to the dynamics of volcanic activity. We show that it is possible to characterize the displacement of the main hydrothermal activity sections generating the SP anomalies measured on the surface in the complex environment of a volcanic edifice. Furthermore, this study supports results from previous work on the behavior of the hydrothermal system within the summit cone of Piton de la Fournaise, characterized by hydrothermal flow along the main active structural limits. For example, the greatest depths obtained from the MWT analyzes during the repose period (1995, Fig. 6) agree with results inferred from a resistivity survey (Lénat et al 2000, 2012a) and correspond to the top of the main hydrothermal reservoir.

Table 3
Depths of the synthetic potential generation area and position calculated by MWT of synthetic SP signals generated by a dipole. The synthetic potential generation area; (Ss) is the reference, (Z) the depth (X) the position used to generate the synthetic SP signal. Cd is the depth calculated by multi-scale wavelet tomography with the wavelets. Solution is the number of calculated depths. Signal/noise ratio (SNR) of infinity and 5 was applied to the synthetic SP signal to investigate noise effects on the calculated depths. Distance along profile (X) and depth (Z) are in m.

| Wavelet type | Dipole | SNR | Solution | Unit (m) | | | |
|---|---|---|---|---|---|---|---|
| | | | | Distance x | σx | Depth z | σz |
| – | Ss | – | – | 1750 to 1810 | – | −100 to −200 | – |
| H2, H3, V2, V3 Cd | | ∞ | 900 | 1779 | 4 | −137 | 8 |
| H2, H3, V2, V3 Cd | | 5 | 400 | 1788 | 15 | −126 | 29 |

Moreover, the depths estimated by MWT delineate the main hydrothermal fluid pathways within the summit cone as shown in Barde-Cabusson et al 2012. The vertical resolution of the MWT depth values allows for quantifying the level of the hydrothermal activity over time with respect to the volcanic activity in the summit cone.

## 7. Conclusion

The hydrothermal system of PdF volcano consists in one extended system with a deep reservoir and several main flow paths rising along the weak edifice structures. Within the summit cone, among the main flow paths, 6 distinct and persistent hydrothermal activity sections were found by this study, at least during the studied period 1993–2008 These active sections spread throughout the summit cone and were associated to pit crater and main faults around the Bory-Dolomieu craters, which are known to have hydrothermal alteration at depth. With the renewed eruptive activity in 1998, the elevation of the hydrothermal activity sections varied by ~400 m from 1998 to 2000 and remained relatively constant at shallow depths until 2008 (end of this study). From 2000, with the resumption of the summit eruption, the hydrothermal activity sections remained at ~2400 m a.s.l., b130 m Below the top of the summit cone (Table 2). This elevation value also correlates well with the fumarole belt observed in the wall of Dolomieu crater following the 2007-collapse event.

This study shows that the application of MWT to time series SP data makes it possible to efficiently and quantitatively determine the depth variation of the hydrothermal activity sections within hydrothermal system, in accordance to the volcanic structure and its plumbing system. Importantly, the 16 years of observations sets a baseline of hydrothermal depth estimates and supports a clear link between the depth of these electrical generation areas and changes in magmatic activity. As such this approach, especially if combined with continuously recording SP systems, can be an important component for monitoring hydrothermal changes that may precede changes in volcanic activity.

## Acknowledgments

This work would not have been possible without the support of the Piton de la Fournaise Observatory (OVPF, IPG Paris) and the University of St Denis, La Réunion, in particular T.  Staudacher and P.  Bachèlery. We thank T.  Spurgeon for his help and discussions and A.  Peltier, L. Letourneur and T.  Dorsch for their help during field work and S. Barde-Cabusson and A. Finizola for the 2006 and 2008 SP data. We also thank CNRS-France for financial support through the National Scientific Programs ACI CatNat and ANR Volcarisk.

Appendix A.  Complementary information on the self-potential
                      process

The electrical potential variations in a volcanic structure are mainly due to the type of electrical generation processes i.e. 1) electrokinetic, 2) thermoelectric processes, and 3) resistivity contrast (e.g. Zlotnicki and Nishida  2003; Lénat  2007; Jouniaux et al 2009 and references therein).

A.1. Electrokinetic processes

The electrokinetic process occurs when water flows through a porous medium. Several parameters will influence the electrokinetic effect including the chemical composition of the flow and mineralogy of the ground and magma. In particular,  the flow pressure removes Ions from the surface of minerals constituting the rock and generates a differential potential between the ions present in solution and the polarized mineral,  which is located in the Helmholtz double layer (Ishido & Mizutani  1981; Avena & de Pauli 1996; Guichet & Zuddas 2003; Hase et al 2003; Aizawa et al 2008; Jouniaux et al 2009). The stronger the water flow or fluid pressure, the stronger the differential ion displacement will be, leading to an increase in the electrical current.

When the electrokinetic effect is the main process, there is an inverse relationship between the topographic surface and the measured electrical anomaly, called the topographic effect (e.g. Corwin & Hoover 1979; Zlotnicki & Nishida 2003; Lénat 2007; Jouniaux et al 2009). The further the top of the water table is from the topographic surface, the more negative will be the electrical potential. On active volcanic systems,  water flows are commonly grouped into two types of environments:  (1) in hydrogeological environments, where the direction of water flow is typically controlled by gravitational forces, the water will generally flow down; (2) in hydrothermal environments, where the fluids are generally rising due to fluid pressure, electrokinetic process will generate a positive SP anomaly.

Self-potential anomalies generated by the electrokinetic effect are typically several tens to several hundreds of mV in amplitude. A related effect, rapid fluid disruption (RFD) (Johnston et al., 2001), is an ephemeral phenomenon which characterises changes in state of the water to vapour phase and will be expressed by an increase of the rising water flux.

A.2. Thermoelectric process

The thermoelectric process occurs when heat flux (e.g., due to magmatic intrusions) is applied to a rock generating a thermal gradient. An increase in temperature will increase the energy of the free ions inside the pore spaces of the rock. The differential displacement of these ions will thus generate an electrical current. Self-potential anomalies generated by thermoelectric processes typically range from a few to several tens of mV in amplitude.

Associated with magmatic intrusion and convective hydrothermal cells, thermal processes also support SP generation, however, past studies have shown that thermal contribution is generally one order of magnitude lower than the electrokinetic process contribution (e.g., heating at the base of the system) (Corwin and Hoover 1979; Zlotnicki and Nishida, 2003).

A.3. Effect of ground resistivity heterogeneity

Another phenomenon is the effect of heterogeneous ground resistivity on electrical potential, which is commonly considered as a secondary effect (Sailhac and Marquis 2001; Saracco et al 2004). However, Minsley et al. (2007) have shown that resistivity contrasts, when not spatially associated with water flow (electrokinetic effect), may be significant in some instances. In this study, however, we investigate a situation where the main resistivity contrasts are mostly linked with underground water flow or hydrothermal systems, which are the main source of  rock alteration (Lénat et al 2012a and references theerin).

A.4. SP/elevation gradient

In order to differentiate from one environment to another, it is common to look at the SP/elevation gradient. The SP/elevation gradient can be symmetrical or asymmetric to the centre of the anomaly, depending on the shape of the topographic surface (Finizola et al., 2002; Lénat, 2007).

Appendix B. Mathematical background of wavelet analyzes

As presented in Mauri et al. (2011), multi-scale wavelet tomography is based on a statistical approach using continuous wavelet analyzes. Here we define a potential generation process to be the process that generates the measured potential field at the surface. The type of potential field is either an electrical field (self-potential), gravity field or magnetic field. Each has specific types of generation processes. Appendix A presents the type of generation processes for the self-potential. The potential generation area is the area at depth where the potential generation is taking place. In order to respect both potential field and wavelet theories, the wavelets used in the analyzes must be from the Poisson kernel family (Poisson operator; Courant and Hilbert, 1962; Nabighian, 1984; Saracco and Tchamitchian, 1990; Blakely, 1995; Moreau et al., 1997, 1999; Saracco et al., 2004, 2007). The general equation of the continuous wavelet transform is presented in several studies (e.g., Grossmann and Morlet, 1984; Saracco, 1987; Saracco et al., 1989, 1990). The dilation, $Da[g(x)]$, and the translation, $T^b[g(x)]$, parameters are defined in nD space as:

$$D^a[g(x)] = a^{-n} g\left(\frac{x}{a}\right) \text{ and } T^b[g(x)] = g(x-b) \qquad (1)$$

Based on Eq. (1), the arbitrary signal $s$ is decomposed into elementary contributions by $g$ called wavelets through dilation $D^a g$, with ($a >$ 0), and translation $T^b g$, in the half-space (b, a) on $\mathbb{R}^n x \mathbb{R}_+^{*\{0\}}$. The general equation is:

$$L(b,a) = T^b D^a \bar{g}, |, s_x = \int a^{-n} s(x) \bar{G}\left(\frac{(x-b)}{a}\right) dx^n \qquad (2)$$

In Fourier space, Eq. (2) becomes:

$$L(b,a) = T^b D^a \bar{g}, |, s_x = \int e^{-ibu} \bar{G}(au) \mid S(u) du^n$$
$$= D^{-a} e^{-ibu} \bar{G}(u) \mid S(u)_u \qquad (3)$$

where $L_{(b,a)}$ is the wavelet coefficient of the signal $s \in R^n$ with the norm $L_1$, decomposed in wavelet contributions dilated and translated in the space (x,z), respectively, from "a" (a < 0) and "b". $g$ is the real or complex admissible analyzing wavelet. G (or S) is the Fourier transform of $g$ (or s), with the admissible condition $G(0) = 0$. $u \in R^n$ is the dual variable in the Fourier domain of $x \in R^n$. $n$ is the order of the derivative of the analyzing wavelet ($g \in R$) with $n \in N$ (Moreau et al., 1997).

As described in the pioneering work of Moreau et al. (1997), the potential field $\Phi(x,z)$ recorded at the summit crater, $\Phi(x,z)$ belongs to $R^2 x R_+^{*\{0\}}$, where $x \in R^2$. The potential field generated by the potential generation process, $\sigma$, verifies $\Delta \Phi(x,z) = -\sigma(x,z)$ in the half-plane (x, z < 0). $\sigma = 0$ for z > 0. When $z \to 0$, $\Phi(x,z) = \Phi_0(x)$. In the Fourier domain, in 2D space, $\hat{\Phi}(u,z)$ verifies Eq. (4):

$$\hat{\Phi}(u,z) = \int D^2 \hat{P}(u) \hat{\Phi}_0(u) du^2 \qquad (4)$$

with $\hat{P}(u) = e^{-2\pi|u|}$ and $\hat{P}(x) = C_{n+1}(1 + |x|^2)^{-(n+1)/2}$. For n = 2, $P(x) = C_3(1 + |x|^2)^{-(3)/2}$.

"$\Phi_0(x)$" is the potential field recorded by our instruments in the horizontal plane (x, z = $z_0$). The Poisson kernel, P, is in $R^2$ and $\hat{P}$ is in the Fourier domain such that: $D^z \hat{P} * D^{z'} \hat{P} = D^{z+z'} \hat{P}$. The dilation operator, D, is acting only on the vertical component, the depth z. $C_{n+1}$ is a constant depending on the dimension order (Courant and Hilbert, 1962; Moreau et al., 1997; Saracco et al., 2007).

Since the depth acts as the dilation parameter for the potential field (2), the kernel Poisson family is analogous to the wavelet family ($\hat{P}(x) = 1$), for which only the derivatives of P are admissible analyzing wavelets (Moreau et al., 1997).

At the surface, z = $z_0$, the measured potential field is generated by the potential generation process, $\sigma$, which is in the half-plane (z < 0).

In 2D horizontal space (i.e., $x \in R^2$), the partial derivative of the Poisson kernel ($\hat{P}$) in Fourier space (Moreau et al., 1997):

$$\frac{\partial^m \hat{P}(u,v)}{\partial u^k \partial v^l} = (-2\pi)^m \left(u^k + v^l\right) e^{-2\pi\sqrt{u^2+v^2}} \qquad (5)$$

with $m = k + l$.

The general equation defines the horizontal and vertical partial derivative of the Poisson Kernel: the $k$th horizontal partial derivative of the Poisson kernel $H_k(u)$ is written (Moreau et al., 1997; Saracco et al., 2004):

$$H_k(u) = \frac{\partial^k \hat{P}(u)}{\partial u^k} = (2\pi i u)^k \exp(-2\pi|u|) \qquad (6)$$

where $u$ is the dual variable in Fourier domain of the spatial variable, x. $k$ the order of the derivative (an integer here), and $i$ the imaginary number, such as $i^2 = -1$.

The general equation of the vertical partial derivative of order $l$ of the Poisson kernel, $V_l(u)$, can be obtained through the Hilbert transform of the horizontal derivative (Moreau et al., 1997; Saracco et al., 2004):

$$V_l(u) = \frac{\partial^l \hat{P}(z)}{\partial z^l} = -HT\left[\frac{\partial^l \hat{P}(u)}{\partial u^l}\right] = -2\pi|u|(2\pi|u|)^{(l-1)} \exp(-2\pi|u|) \qquad (7)$$

with $u$ the dual variable of variable x. $l$ is the order of the derivative, such that $l \in N$, and $i$ is the imaginary number. HT is the Hilbert transform.

As presented in the Eq. (1), the results from the wavelet analysis of a signal, s, by a wavelet, g, at the dilation, a, along the entire signal is $L_{(b,a)}$. When performing continuous wavelet analyzes (CWT), the wavelet analyzis of the signal, s, is not only done for one dilation value, but rather over a range of dilation values (Grossmann and Morlet, 1984; Saracco et al., 1989; Saracco, 1994; Moreau et al., 1997). Therefore, over a range of dilation [$z_{min}$; $z_{max}$], the result of the CWT is a matrix made of coefficient correlations, where each row of the matrix represents how good the correlation is between the analyzing wavelet based on the Poisson Kernel and the measured potential field at a specific altitude $z_0$.

If the reconstructed potential field over a range of dilation (i.e., a range of altitudes) is made of at least one anomaly generated by a potential generation process (i.e., self-potential) at depth, then the matrix of coefficient correlations will present both minima and maxima of coefficient correlations, which will be organized in lines converging toward the potential generation area (Fig. 7c). These lines, consisting of either minima or maxima of coefficient correlations, are called extrema. These lines are not a priori straight lines, but rather oscillating lines due to the heterogeneous property of the volcanic structure (i.e., porous structure), as seen Fig. 3.

The depth of the potential generation area can then be estimated using the properties of the potential field theories, which link the potential generation process to the measured potential field, roughly called the anomaly. Through a linear regression of each extrema, it then becomes possible to estimate their intersection (for more detail on the mathematical theory sees the work of Moreau et al., 1997). Eq. (8) presents the mathematical relationship between dilation $l$ and depth Z, as presented in the MWT code (Mauri et al., 2011, additional electronic supplement).

$$Z = 2 \lambda p l \qquad (8)$$

The sampling step is $p$, and the length of the support base of the wavelet, $\lambda$, is predefined in the module of the wavelet definition source code to ensure that the wavelet is always well defined and is unitless (Mauri et al., 2011, additional electronic supplement).

Our method is based on a statistical approach by using several wavelet analyzes. This allows us to define a domain of validity for each depth of potential generation area. Two synthetic examples of MWT applied on synthetic SP signals are presented in Figs. 7 and 8. Each synthetic example presents result of signals with and without noise. The results in Fig. 7 are presented in Table 3 and show the case of a flat topography with a vertical low resistivity body at depth that surrounds a non-punctual dipole. The other example (Fig. 8), presents the case of a topography with a 10% slope and an asymmetric low resistivity body at depth that surrounds a dipole.

Results from the different examples show that noise on SP signal has a limited impact on the depth determination through the statistical approach of the MWT method. Indeed, the noise acts as a high frequency signal in comparison to the signal frequency of the SP anomaly. Therefore, by using dilation of higher values (in the case of Fig. 7, dilation ≥ 4), it becomes possible to reduce the effect of the noise on the depth calculation.